\documentstyle[12pt]{article}
\begin{document}
\thispagestyle{empty}
\begin{center}{THE SL(2,C) GAUGE THEORY OF GRAVITATION AND THE QUANTIZATION OF THE
GRAVITATIONAL FIELD}\end{center}
\vspace{1cm}
\begin{center}{Moshe Carmeli}\end{center} 
\begin{center}{Department of Physics,}\end{center} 
\begin{center}{Ben Gurion University, Beer Sheva 84105, Israel}\end{center}
\begin{center}{and}\end{center}
\begin{center}{Shimon Malin}\end{center}
\begin{center}{Department of Physics and Astronomy,}\end{center} 
\begin{center}{Colgate University, Hamilton, NY 13346, U. S. A.}\end{center}
\vspace{1cm}
\begin{center}{\bf Abstract}\end{center}
A new approach to quantize the gavitational field is presented. It is based on
the observation that the quantum character of matter becomes 
more significant as one gets closer to the Big Bang. As the metric looses its
meaning, it makes sense to consider Schr\"{o}dinger's three generic types of 
manifolds -- unconnected differentiable, affinely connected and metrically
connected -- as a temporal sequence following the Big Bang. Hence one should 
quantize the gravitational field on general differentiable manifolds or on
affinely connected manifolds. The SL(2,C) gauge theory of gravitation is
employed to explore this possibility. Within this framework, the quantization
itself may well be canonical.     
\newpage
\section{Introduction}
Why it is impossible to quantize the gravitational field in the same way that
other fields, such as the EM field, are quantized? The answer to this question 
is well-known: The standard quantization procedure applies to fields that are
defined over a spacetime of a given metric structure. In the case of the
gravitational field, however, there is no metric structure given apriori. It 
is, in fact, this very metric structure that functions as the gravitational 
field.

The response to this difficulty spans a whole spectrum, from a ``minimal" to
a ``maximal" approach. The ``minimal" approach assumes a background of a flat
metric structure,
$$g_{\mu\nu}=\eta_{\mu\nu}(\mbox{\rm flat})+h_{\mu\nu},\eqno(1)$$
and treats the deviations $h_{\mu\nu}$ from flatness as the field to be 
quantized. In the ``maximal" approach one tries for quantization in the
context of grand unification.
\section{The Present Proposal} 
We take the middle ground -- it is a way to fully quantize the gravitational
field, in all its non-linearity; but the suggested quantization procedure 
applies to the gravitational field only.

Consider this: The closer we get to the Big Bang, the more relevant are the
quantum features. On the other hand, as we get down to the times of the order
of Planck's time, metric time and space lose their meaning.

This last statement is based on the following arguments: In general, matter as
we know it does not exist at Planck's time, so we don't have an operational 
definition of metric relationships.

More specifically, Amati {\it et al.} [1] and Konishi {\it et al.} [2] have 
found, in the context of string theory, that there is a minimal observable 
length. A similar result was obtained on more general ground by 
N. Itzhaki [3]. This leads
to the following question: Is there a cosmological phase which is, generically
as well as temporally, prior to metric spacetime, and is it possible to 
quantize gravity at this phase?

Note that if the answer is ``yes", we will have a way of resolving the problem
that was mentioned at the beginning: The fact that in the case of the 
gravitational field, unlike the electromagnetic and other fields, we don't
have a spacetime of a definite structure on which the field is defined.

The thrust of the present approach is to show that the answer is, indeed, 
``yes": {\it There is a cosmological phase which is generically prior to
metric spacetime, and it should be possible to quantize gravity at this phase.}

Mathematically, this possibility is based on the SL(2,C) gauge theory of 
gravitation, as we shall see.
\section{A Classification of Manifolds}
Consider the following three generic phases leading to metric spacetimes [4]:
\newline
(1) The general differentiable manifold (unconnected),\newline
(2) Affinely connected manifold,\newline
(3) Metrically connected manifold.

There is an intermediate phase between (2) and (3): Spacetime with a 
rudimentary metric structure (STRMS): In metrically connected spacetimes
$$\Gamma^{\alpha}_{\hspace{5pt}\mu\alpha}=\left(\ln\sqrt{-g}\right)_{,\mu}
\equiv\phi_{,\mu}.\eqno(2)$$
Hence it is possible, in nonmetric, affinely connected spacetimes, to define
$\phi$ as the potential of $\Gamma^{\mu}_{\hspace{5pt}\mu\alpha}$, and let
$e^{\phi}$ play the role of $\sqrt{-g}$. Equation (2) determines the volume element
up to an overall multiplicative factor. STRMS is metric only in the sense of
having a volume element; $ds^2$ is not defined in it.

Perhaps surprisingly, a lot can be done on general differentiable manifolds.
Covariant, contravariant and mixed tensors, tensor densities and spinors can be
defined. (However, the usual relationship between covariant and contravariant 
components does not hold -- there is no metric tensor to connect them!); there
are invariant integrals which are obtained by integrating over tensor 
densities: If $\pounds$ is a density, then
$$\int\pounds d^4x\eqno(3)$$
is invariant, and hence we can have a Lagrangian formalism, with 
Euler-Lagrange equations.
\section{The SL(2,C) Gauge Theory of Gravitation}
The theory evolved out of The Newman-Penrose null tetrad formalism [5]. The 
latter involves, at each point of spacetime, a tetrad of null vectors, 
$l_\mu$ and $n_\mu$ (real), $m_\mu$ and $\overline{m}_\mu$ (complex), which 
satisfy
$$l_\mu n^\mu=-m_\mu\overline{m}^\mu=1,\eqno(4)$$
the tetrad components of the Weyl and Ricci tensors, and the spin coefficients. 
For example, the complex tetrad components of the Weyl tensor are
$$\psi_0=-C_{\mu\nu\rho\sigma}l^\mu m^\nu l^\rho m^\sigma$$
$$\psi_1=-C_{\mu\nu\rho\sigma}l^\mu n^\nu l^\rho m^\sigma$$
$$\psi_2=-\frac{1}{2}C_{\mu\nu\rho\sigma}\left(l^\mu n^\nu l^\rho n^\sigma-
l^\mu n^\nu m^\rho\overline{m}^\sigma\right),\eqno(5)$$
$$\psi_3=-C_{\mu\nu\rho\sigma}\overline{m}^\mu n^\nu l^\rho n^\sigma$$
$$\psi_4=-C_{\mu\nu\rho\sigma}\overline{m}^\mu n^\nu\overline{m}^\rho n^\sigma$$

Carmeli's SL(2,C) gauge theory of gravitation [6] is a group theoretical
formulation of the Newman-Penrose formalism. Both formalisms are equivalent 
to General Relativity, but the SL(2,C) gauge theory of gravitation can be
formulated on affinely connected manifolds [7] and, as we found in the context
of the present work, the SL(2,C) gauge theory of gravitation can be formulated
even on general differentiable manifolds.
\section{The Structure of the SL(2,C) Gauge Theory of Gravitation}
Start out with a {\it spin frame} defined at each point of a manifold: Two
linearly independent 2-component spinors satisfying 
$$l_An^A=1.\eqno(6)$$
(Raising and lowering indices is done with the standard antisymmetric spinor
matrix.) The components of $l_A$ and $n_A$ are expressed as the components of 
a 2$\times$2 matrix $\zeta$. It follows from Eq. (6) that the matrix $S$
relating any two spin frames,
$$\zeta'=\left(\begin{array}{c}l_A'\\n_A'\\\end{array}\right)=S^{-1}\left(
\begin{array}{c}l_A\\n_A\\\end{array}\right)=S^{-1}\zeta,\eqno(7)$$
belongs to the group SL(2,C).

The introduction of local gauge invariance under the transformations $S$ of
the spin frame brings about a (compensating) {\it gauge field} of as follows:
Since, in general,
$$\nabla_\mu\left(S^{-1}\zeta\right)\neq S^{-1}\nabla_\mu\zeta,\eqno(8)$$
we introduce a gauge vector field 2$\times$2 matrices $B_\mu$ that transform 
according to
$$B_\mu'=S^{-1}B_\mu S-S^{-1}\partial_\mu S, \eqno(9)$$
and then
$$\left(\nabla_\mu-B_\mu'\right)\left(S^{-1}\zeta'\right)=
S^{-1}\left(\nabla_\mu-B_\mu\right)\zeta.\eqno(10)$$
In analogy with the Yang-Mills field these are called {\it potentials}.

The corresponding {\it fields} are defined, also in analogy with Yang-Mills
theory, as
$$F_{\mu\nu}=\partial_\nu B_\mu-\partial_\mu B_\nu+\left[B_\mu,B_\nu\right],
\eqno(11)$$
where
$$\left[B_\mu,B_\nu\right]=B_\mu B_\nu-B_\nu B_\mu.\eqno(12)$$
The fields transform as follows:
$$F_{\mu\nu}'=S^{-1}F_{\mu\nu}S.\eqno(13)$$
\section{The Lagrangian}
There are a number of equivalent ways to write down the Lagrangian density in
the context of metric spacetimes. One of these ways can be used in general
differentiable manifolds:
$$\pounds=-\frac{1}{2}\mbox{\rm Tr}\left\{\epsilon^{\alpha\beta\gamma\delta}
F_{\alpha\beta}\left(-\frac{1}{2}F_{\gamma\delta}+B_{\gamma,\delta}-
B_{\delta,\gamma}+\left[B_\mu,B_\nu\right]\right)\right\}.\eqno(14)$$
The potentials and fields are considered independent for the purpose of the
variational procedure. The Euler-Lagrange equations are:
$$F_{\mu\nu}=B_{\mu,\nu}-B_{\nu,\mu}+\left[B_\mu,B_\nu\right],\eqno(15)$$
$$\epsilon^{\alpha\beta\gamma\delta}\left\{\partial_\delta F_{\alpha\beta}-
\left[B_\delta,F_{\alpha\beta}\right]\right\}=0.\eqno(16)$$ 
In Eqs. (14)-(16) only covariant components of the fields and potentials 
appear. Hence the absence of correspondence between covariant and 
contravariant components in non-metric manifolds is not a problem.
\section{Quantization}
The straightforward approach is to quantize Eq. (16) by taking the usual 
commutation relations between the matrix elements of the $B_\mu$ and their 
canonical conjugates. It is not clear, however, how to resolve the problem of
redundant components, i.e., the existance of more functions than the number of 
degrees of freedom. There is, however, a different approach, one that worked
very well for the linearized equations of General Relativity in the 
Newman-Penrose formalism [8]:

The sets of variables of Newman and Penrose are equivalent to the $B_\mu$ and
the $F_{\mu\nu}$'s. In the case of the linearized approximation to the 
Newman-Penrose formalism, it turns out that all the functions can be expressed
in terms of one complex function, $\psi_2$. The problem of quantization 
reduces, then, to quantizing one complex function. This is done as follows.

Expanding in the Wigner matrix elements $D^j_{\hspace{5pt}sm}$ of the 
irreducible representations of the group SU(2):
$$\psi_2\left(t,r,\theta,\phi\right)=\sum_{j=0}^\infty\sum_{m=-j}^j
\alpha_{\hspace{5pt}2m}^j\left(t,r\right)D^j_{\hspace{5pt}0m}
\left(\theta,\phi\right).\eqno(17)$$
Using the Newman-Penrose equations one (eventually) gets a separate partial
differential equation for each of the $\alpha_{\hspace{5pt}2m}^j$'s: 
$$\left[3\frac{\partial^2}{\partial t^2}+2\frac{\partial^2}{\partial t
\partial r}-\frac{\partial^2}{\partial r^2}+\frac{j\left(j+1\right)}{r^2}
\right]\left(r^2\alpha_{\hspace{5pt}2m}^j\right)=0,\eqno(18)$$
so that the $\alpha_{\hspace{5pt}2m}^j$ can be taken as annihilation 
operators, and the standard commutation relations between $\alpha_{\hspace{5pt}2m}^j$
and $\overline{\alpha}_{\hspace{5pt}2m}^j$ can be postulated. The 
generalization of this approach to the non-linear case is mathematically 
challenging; there seems to be no reason of principle, however, that prevents
it from being carried out. 
\section{Summary} 
The following are the key elements of the present approach:\newline
1. Since the quantum characteristics of matter get more significant as one
gets closer to the Big Bang, while the metric characteristics loose their 
meaning, it makes sense to consider Schr\"{o}dinger's three generic types of 
manifolds (unconnected differentiable, affinely connected and metrically
connected) as a temporal sequence following the Big Bang. (The word 
``temporal" is used here in the sense of time sequence, not time measurement.)
\vspace{2mm}\newline
2. For the same reason it makes sense to try for quantization of the 
gravitational field on general differentiable manifolds (if possible) or on
affinely connected manifolds, with or without rudimentary metric structure.\vspace{2mm}\newline
3. The SL(2,C) gauge theory of gravity seems to be the best existing theory 
for the exploration of this possibility.\vspace{2mm}\newline
4. Within this framework, the quantization itself may well be canonical. 
\newpage
\section*{References}
1. D. Amati, M. Ciafaloni and G. Veneziano, {\it Phys. Lett.} {\bf B216}, 41 
(1989).\newline
2. K. Konishi, G. Paffuti and P. Provero, {\it Phys. Lett.} {\bf B234}, 276 
(1990).\newline
3. N. Itzhaki, private communication with S. M.\newline
4. E. Schr\"{o}dinger, {\it Space-Time Structure} (Cambridge Univ. Press,
1954).\newline
5. M. Carmeli, {\it Classical Fields, General Relativity and Gauge Fields}
(John Wiley, 1982), Chap. 3.\newline
6. {\it ibid.}, Chap. 8.\newline
7. M. Carmeli and S. Malin, {\it Ann. Phys.} (NY) {\bf 103}, 208 (1977).\newline
8. S. Malin, {\it Phys. Rev.} {\bf D10}, 2238 (1974).
\end{document}